# Mixing order asymmetry in nanoparticle-polymer complexation and precipitation revealed by isothermal titration calorimetry

**Letícia Vitorazi[1,2,3] and Jean-François Berret[1*]**

[1]Université Paris Cité, CNRS, Matière et systèmes complexes, 75013 Paris, France
[2] Laboratório de Polímeros, Nanomateriais e Química Supramolecular, EEIMVR, Universidade Federal Fluminense, Avenida dos Trabalhadores, 420, Volta Redonda RJ, CEP 27225-125, Brazil.
[3]Programa de Pós-Graduação em Engenharia Metalúrgica, EEIMVR, Universidade Federal Fluminense, Avenida dos Trabalhadores, 420, Volta Redonda RJ, CEP 27225-125, Brazil.

**Abstract:** In recent years, there has been a renewed interest in complex coacervation, driven by concerted efforts to offer novel experimental and theoretical insights into electrostatic charge-induced association. While previous studies have primarily focused on polyelectrolytes, proteins or surfactants, our work explores the potential of using cerium ($CeO_2$) and iron ($\gamma$-$Fe_2O_3$) oxide nanoparticles (NPs) to develop innovative nanomaterials. By combining various charged species, such as polyelectrolytes, charged neutral block copolymers and coated NPs, we study a wide variety of complexation patterns and compare them using isothermal titration calorimetry, light scattering and microscopy. These techniques confirm that the titration of oppositely charged species occurs in two steps: the formation of polyelectrolyte complexes and subsequent phase (or microphase) separation, depending on the system studied. Across all examined cases, the entropic contribution to the total free energy surpasses the enthalpic contribution, in agreement with counterion release mechanisms. Furthermore, our investigation reveals a consistent asymmetry in the reaction enthalpy associated with the secondary process, with exothermic profiles observed upon the addition of cationic species to anionic ones and endothermic profiles in the reverse case.

# I – Introduction

Complex coacervation is a liquid-liquid phase separation that occurs in solutions of oppositely charged colloidal species, such as proteins, polymers and surfactants. At the transition, oppositely charged species spontaneously associate to form a dense coacervate phase in equilibrium with a supernatant. This phenomenon was initially identified more than a century ago and later examined in details in mixtures of proteins and polysaccharides by Bungenberg de Jong.[1] In recent years, complex coacervation has undergone a renewed interest thanks to combined efforts to obtain precise experimental and theoretical accounts of the phenomenon, describing in details the phase diagrams, the role of ionic strength, the surface tension or the rheology of the coacervate phase.[2-13] A variant of the coacervation transition has also been noted in select systems that show liquid-solid phase coexistence instead, with the solid phase referred to as a complex precipitate.[14-17] In other well-studied cases, when polyelectrolytes are replaced by charge-neutral block copolymers, the electrostatic association leads to a microphase separation, and to the spontaneous formation of complex coacervate core micelles.[14,18-27] The driving force for association has been found to be of enthalpic and entropic origin, as revealed by isothermal titration calorimetry (ITC).[16,27-33] The enthalpic part of the association free energy is linked to the pairing of opposite charges, while the entropic contribution to the free energy comes from the release of counterions that are originally condensed on the colloid surface or along the polymer backbone.

In the field of complex coacervation, the most extensively studied systems involve oppositely charged synthetic polymers with matching degrees of polymerization and charge.[3,6,8,18] Far from charge stoichiometry however, co-assembly also takes place, giving rise to colloids made of entangled and loosely bound polymers, known as polyelectrolyte complexes (PECs).[3,18,34-36] Because their formation depends on physicochemical conditions such as concentration, pH and ionic strength and also the mixing mode, it is assumed that PECs are out of equilibrium, and in a metastable state.[3,36] Apart from the species mentioned above - polymers, surfactants and proteins, most charged molecular and colloidal systems, including multivalent ions,[26,37,38] lipids assemblies,[39,40] biological polymers,[19,33,41-44] nanoparticles,[45,46] cellulose nanofibers[47,48] nanoplastics,[49] form electrostatic complexes by association with oppositely charged macromolecules, by analogy with PECs. In many of the previous instances, co-assembled structures exhibit endothermic signatures similar to those of coacervation and precipitation, and are typically characterized by entropy-driven mechanisms.[19,26,42,44,48]

Another well-studied class of co-assembled electrostatic complexes involves inorganic nanoparticles (NPs) made from metal oxide or noble metal with oppositely charged polymers for non-covalent functionalization. In this process, known as the co-assembly grafting-to technique,[5,50,51] polymer-NP complexes form through the spontaneous adsorption of functional polymers onto the NP surface. This method produces hybrid core-corona structures, where the charged or neutral polymer corona offsets van der Waals interactions, preventing NP aggregation in complex and physiological environments and expanding their potential applications. Despite the extensive research on NP functionalization, there has been however limited investigation into the polymer-induced coacervation or precipitation of inorganic NPs to date.[2,6] In this work, we study the phase behavior and thermodynamics of the electrostatic complexation of sub-10 nm cerium ($CeO_2$) and iron ($\gamma$-Fe$_2$O$_3$) oxide NPs with oppositely charged polymers. Both NPs are coated with negatively charged, low-molecular-weight poly(acrylic) acid chains. These particles were selected due to their wide range of potential applications in nanomedicine. Specifically, CeO$_2$ NPs are known as reactive oxygen species scavengers, while $\gamma$-Fe$_2$O$_3$ serves as a contrast agent for magnetic resonance imaging.[50] As cationic polymers, poly(diallyldimethylammonium chloride) homopolymers and the charged-neutral block copolymers poly(trimethylammonium ethylacrylate methyl sulfate)-b-poly(acrylamide) are used.[14,30,52] The enthalpy, binding constant, and stoichiometry of the polymer-NP reactions were determined by ITC as a function of the charge ratio and the mixing order. For the four pairs of oppositely charged systems, *i.e.* polymer/polymer, polymer/NP, copolymer/polymer and copolymer/NP, the titration reaction takes place in two successive steps: the primary process is the formation of charged polyelectrolyte complexes whereas the secondary process is the transition towards a coacervate, precipitate or microphase separation. Moreover, we observe a systematic asymmetry with respect to the mixing order for the secondary process: it is exothermic upon polycation addition and endothermic in the reverse case.

## 2 – Materials and Methods

### 2.1 – Materials

Poly(acrylic acid) in sodium salt form, hereinafter abbreviated PAA ($M_n$ = 3.0 kDa, Đ = 1.8) and poly(diallyldimethylammonium chloride), abbreviated PDADMAC ($M_n$ = 7.6 kDa, Đ = 3.5)[52] were purchased from Sigma-Aldrich. Sodium hydroxide (NaOH) was supplied by Riedel-de Haën and 28-30% ammonia solution by Technic France. The block copolymer poly(trimethylammonium ethylacrylate methyl sulfate)-b-poly(acrylamide), referred to as PTEA-b-PAm and cerium dioxide (CeO$_2$) NPs were provided to us by Solvay, Centre de Recherche d'Aubervilliers (Aubervilliers, France). PTEA-b-PAm block copolymer was synthesized by controlled radical polymerization, MADIX®.[21,53] The number average molar masses for the PTEA and PAm blocks targeted by the synthesis were respectively 11 kDa and 30 kDa. The iron oxide NPs ($\gamma$-Fe$_2$O$_3$, maghemite) were kindly donated by the PHENIX laboratory at Sorbonne University (France). All chemicals were used as received, without further purification. Ultrapure water with a resistivity of 18.2 MΩ cm$^{-1}$ was obtained from a MilliQ® purification system. To ensure the complete deprotonation of PAA weak polyelectrolyte, dispersions were prepared at pH 10.[30] Fig. 1a shows the representation of chemical structure of the polymers. The number of monomers for each of the polymer chains studied, PAA, PDADMAC, PTEA and PAm are 32; 50, 41 and 422 respectively.

### 2.2 – Methods

#### 2.2.1 - Nanoparticles synthesis

The CeO$_2$ dispersion was synthesized *via* thermohydrolysis of an acidic solution containing cerium-IV nitrate salt, Ce(NO$_3$)$_4$, resulting in the homogeneous precipitation of cerium oxide nanocrystals at pH 1.4.[54-56] The CeO$_2$ NPs were shown to exhibit a cubic structure, with a face-centered cubic lattice parameter of 0.5415 nm. High-resolution transmission electron microscopy revealed that the NPs consisted of isotropic agglomerates of 4 to 5 tightly bound 2.5 nm crystallites, leading to an average size of 7.8 nm.[56] The $\gamma$-Fe$_2$O$_3$ dispersion was synthesized following the Massart method *via* the coprecipitation of iron(II) and iron(III) ions induced by the addition of concentrated ammonium hydroxide. This process was followed by oxidation using ferric nitrate, leading to $\gamma$-Fe$_2$O$_3$ maghemite.[57] $\gamma$-Fe$_2$O$_3$ NPs were then sorted with respect to their sizes using successive phase-separation.[58] At the end of the synthesis, the samples manifest as highly concentrated dispersions, with a concentration of approximately 10 wt. % and a pH of 1.8. Measurements from transmission electron microscopy and vibrating sample magnetometry indicated consistent size distribution with median diameter of 6.8 nm and a dispersity index of 0.18. Transmission electron microscopy images of the CeO$_2$ and $\gamma$-Fe$_2$O$_3$ NPs are shown in Supporting Information S1.

**2.2.2 - Nanoparticle coating**
As synthesized, uncoated CeO$_2$ and $\gamma$-Fe$_2$O$_3$ dispersions are stabilized via electrostatic repulsive forces in acidic condition. As pH or ionic strength increase, the NP dispersions destabilize, leading to irreversible aggregation. For applications under physiological conditions, a poly(acrylic acid) coating has been implemented via the precipitation-redispersion method.[55] To this aim, a dilute NP dispersion (concentration 1 g L$^{-1}$, pH 1.5) was added dropwise under vigorous stirring to a PAA solution at the same pH and concentration. The mixed dispersion rapidly became cloudy and precipitated. It was then centrifuged to separate the aggregated NPs from the supernatant. The pH of the precipitate was then raised to pH 10 by addition of NH$_4$OH solution, leading to complete redispersion of the CeO$_2$ or $\gamma$-Fe$_2$O$_3$ NPs. This step was followed by dialysis against DI water using a 10 kDa dialysis membrane (Spectra/Por ®6) overnight to remove the excess polymer chains. Using dynamic light scattering, the hydrodynamic diameter of the CeO$_2$@PAA and $\gamma$-Fe$_2$O$_3$@PAA was found to increase by 5 nm for both CeO$_2$ and $\gamma$-Fe$_2$O$_3$ NPs, indicating the presence of a polymer layer 2.5 nm thick.[21,36] In our previous research, we demonstrated that both PAA coated NPs exhibit remarkable colloidal stability under various conditions, including variations in concentration, pH, ionic strength or plasma proteins.[36,50,55] In this study, we found that complexation of NPs with polycations does not lead to detachment of PAA chains from the particle surface. In Supporting Information S2 we derive the conversion factor $A$ between the weight concentration of non-coated particles, $c_{NP}$ and the molar concentration of negative charges $[-]$ carried by these NPs. We found $A = 4.1 \times 10^{-3}$ mol g$^{-1}$ and $5.5 \times 10^{-3}$ mol g$^{-1}$ for CeO$_2$@PAA and $\gamma$-Fe$_2$O$_3$@PAA NPs respectively

**2.2.3 - Mixing protocol**
Nanoparticle and polymer solutions underwent mixing through two distinct protocols: direct mixing or titration. For direct mixing, cationic stock polymer and anionic NP dispersions, prepared at the same charge molar concentration, noted $[+]$ and $[-]$ respectively were mixed at pH 10 in a single step. The injected volumes were adjusted so that the charge ratios $Z_{+/-}$ or $Z_{-/+}$ varied between 10$^{-2}$ to 100. In the titration protocol, we followed the ITC procedure outlined in the subsequent section. Both protocols, direct mixing and titration, were conducted in two different addition orders. In Type I mixing, the cationic polymers were incrementally added to the NP dispersions while in Type II, the process was reversed, with anionic-coated NPs progressively added to cationic polymer dispersions. This distinction is based on our previous findings indicating asymmetry in thermodynamic titration curves depending on the mixing order.[30] The schematic representation in Fig. 1b illustrates the experimental set-up.

**2.2.4 - Isothermal titration calorimetry (ITC) measurements and data treatment**

Binding isotherms were acquired using a Microcal VP-ITC calorimeter (Northampton, MA) with a cell volume of 1.4643 mL, operating at 25°C with an agitation speed of 307 rpm. Prior to experimentation, samples underwent degassing through the ThermoVac system (MicroCal) using simultaneous vacuum and stirring on a magnetic stir plate for 5 minutes. In a typical run, the syringe was loaded with a NP or polymer solution, ten times more concentrated than that in the measuring cell. The reference cell was filled with degassed ultrapure water. The titration protocol included a preliminary injection of 2 µL, followed by 28 injections of 10 µL at 10 to 20-minute intervals. The standard ITC outcome includes a thermogram depicting the differential power needed to maintain the cell and reference temperatures identical, and a binding isotherm. Calibration of the Microcal VP-ITC calorimeter was carried out on a monthly basis using calcium chloride ($CaCl_2$, 1 mM) and 0.1 mM ethylenediaminetetraacetic acid (EDTA) solution and MES (2-(N-morpholino)ethanesulfonic acid), 10 mM) buffer (Supporting Information S3). The binding isotherm shown in Fig. 1c features the incremental addition of a $[-]$ = 5 mM $CeO_2$@PAA dispersion to a $[+]$ = 0.5 mM PDADMAC solution. Next, integration of the differential power as a function of time for each injection provided the quantity ΔH expressed in kJ mol$^{-1}$, representing the quantity of heat exchanged during the reaction, normalized to ligand concentration. Throughout the ITC measurement campaign, in addition to the calibration already mentioned, PAA, PDADMAC, PTEA-b-PAm, CeO2@PAA and γ-$Fe_2O_3$@PAA dispersions were systematically assessed to determine the heat exchanges associated with their dilution in DI-water at pH 10. These dilution data were then subtracted from the ΔH -measurements to give the net thermodynamic response associated to the complexation reactions. Examples of data processing are provided in Supporting Information S4. Subsequently, the concentrations of titrant and titrate were converted to charge ratios, $Z_{+/-}$ or $Z_{-/+}$, depending on whether the species added during titration were cationic or anionic, respectively.

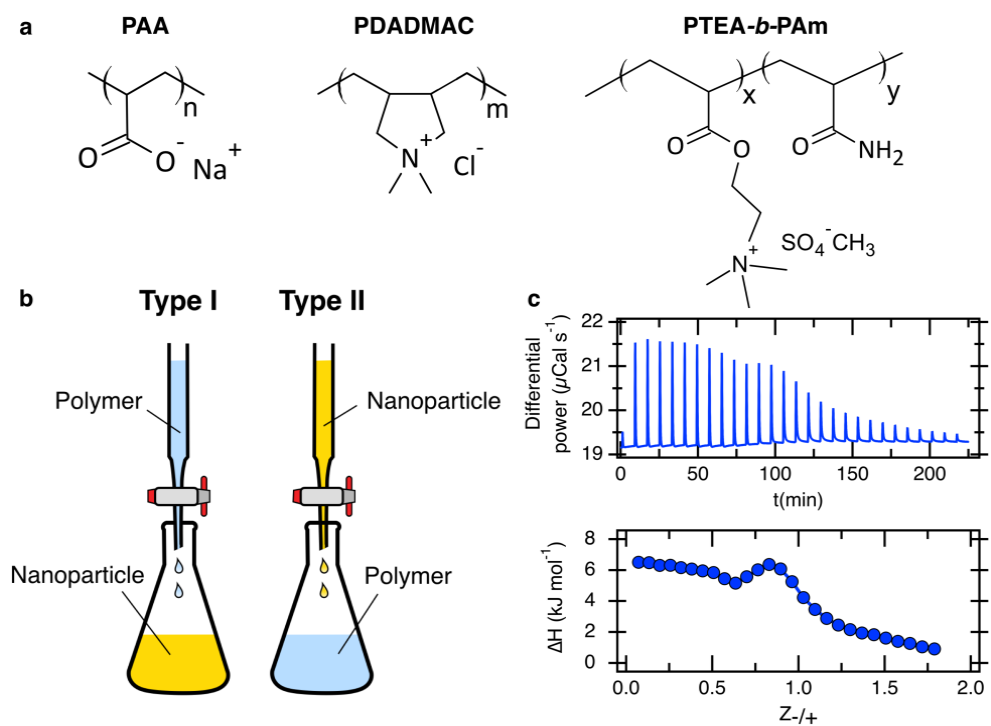

**Figure 1**: Chemical structures and abbreviations of the polymers studied in this work. In a), PAA stands for poly(sodium acrylate), PDADMAC for poly(diallyldimethylammonium chloride) and PTEA-*b*-PAm for poly(trimethylammonium ethylacrylate methyl sulfate)-*b*-poly(acrylamide). b) Schematics of Type I and Type II mixing. c) Example of thermogram showing the differential power *versus* time (upper panel) and binding isotherm *versus* charge ratio (lower panel) obtained for titrating a 0.5 mM PDADMAC solution with a 5 mM CeO$_2$@PAA dispersion.

### 2.2.5 - Light scattering and electrophoretic mobility

Light scattering and $\zeta$-potential analyses for the polymers and coated NPs were conducted using a NanoZS Zetasizer (Malvern Panalytical). The automatic titrator (Malvern Panalytical, MPT-2), synchronized with the light scattering and electrophoresis set-ups, facilitated time-lapse experiments with 10 μL injections every 10 minutes, employing both Type I and II titration methods for structure and charge measurements. In light scattering (173° detection angle), variations in hydrodynamic diameter $D_H(t)$ and scattering intensity $I_S(t)$ over time were recorded, providing insights into the aggregation state of the dispersions. In the Malvern Panalytical software that operates the spectrometer, $I_S(t)$ is the derived count rate (DCR), expressed in kcps. The time axis was later translated into charge ratios, yielding to $D_H(Z_{+/-})$ and scattering intensity $I_S(Z_{+/-})$ for Type I titration, and $D_H(Z_{-/+})$ and scattering intensity $I_S(Z_{-/+})$. The hydrodynamic diameter $D_H$ of the colloids was determined from the diffusion constant using the Stokes-Einstein relation. The interpretation of autocorrelation functions from scattered light employed both the cumulant method and the CONTIN fitting procedure available from the instrument software. Phase analysis light scattering (at a 16° detection angle) complemented the study, enabling the determination of electrophoretic mobility $\mu_E$, subsequently yielding $\zeta$-potential. Experimental conditions, including concentration, pH, and temperature, closely matched those of the ITC set-up.

### 2.2.6 - Optical microscopy

Bright-field and phase-contrast images of coacervate phases in PDAMAC/CeO$_2$@PAA dispersions were captured employing an IX73 inverted microscope (Olympus) equipped with 40× and 60× objectives. Data acquisition and processing were monitored using an Exi Blue camera (QImaging) and managed through Metamorph (Universal Imaging Inc.) and ImageJ softwares.[59] Image processing techniques were applied to assess the size distributions of the coacervate droplets.

### 2.2.7 - Cryo-transmission electron microscopy

Cryo-transmission electron microscopy (cryo-TEM) investigations were conducted using PTEA-*b*-PAm/γ-Fe$_2$O$_3$@PAA dispersions prepared at charge ratio Z = 1 and a concentration of c = 2 g L$^{-1}$. In the cryo-TEM procedure, a minimal volume of the solution was deposited onto a TEM-grid coated with a 100 nm-thick polymer perforated membrane. After carefully removing excess solution via blotting with filter paper, the grid was rapidly immersed in liquid ethane to achieve prompt freezing, preventing crystallization of the aqueous phase. Following this, the membrane-loaded grid was introduced into the vacuum column of a TEM microscope (JEOL 1200 EX operating at 120 kV), where it was kept at liquid nitrogen temperature. Cryo-TEM experiments were performed at a magnification of 40000×.

**2.2.8 - ITC data analysis**

The ITC data analysis was performed via a modified version of the Multiple Non-Interacting Sites (MNIS) model. The model assumes that the entities undergoing titration, known as "macromolecules", possess multiple binding sites to which "ligands" can bind. Moreover, the binding probability is not affected by the occupancy status of other sites on the same macromolecule. It is important to highlight that using both Type I and Type II titrations, cationic polymers and anionic NPs interchangeably assume the roles of "ligands" and "macromolecules." The parameters that describe the reaction are $\Delta H_b$ the binding enthalpy, $K_b$ the binding constant and $n$ the reaction stoichiometry. $n$ denotes the number of non-interacting binding sites available on each macromolecule. Introducing the electrostatic charge ratio $Z$ in place of the polymer and NP concentrations, the enthalpy obtained from thermograms reads:[24,60,61]

$$\Delta H(Z, n, r) = \frac{1}{2}\Delta H_b \left(1 + \frac{n - Z - r}{\sqrt{(n + Z + r)^2 - 4Zn}}\right) \quad (1)$$

where $r = 1/K_b[M]$ and $[M]$ the molar concentration of macromolecules. In this work, we demonstrate that binding enthalpy curves exhibit a sequence of two successive reactions during the titration process. Based on the MNIS model (Eq. 1), we assume that the ITC data obtained here can be accounted for by an expression of the form:

$$\Delta H(Z) = \Delta H^P(Z, n_P, r_P) + \alpha(Z)\Delta H^S(Z, n_S, r_S) \quad (2)$$

where $\Delta H^P(Z, n_P, r_P)$ and $\Delta H^S(Z, n_S, r_S)$ are the enthalpic contributions for the primary and secondary processes, respectively. As a result, the binding enthalpies of each process are noted $\Delta H_b^P$ and $\Delta H_b^S$, and the stoichiometry coefficient $n_P$ and $n_S$. The coefficients $r_P$ and $r_S$ are related to the binding constants $K_b^P$ and $K_b^S$ through the expressions: $r_P = 1/K_b^P[M]$ and $r_S = 1/K_b^S[M]$. The function $\alpha(Z)$ in Eq.2 is there to ensure the continuous transition between the two observed reactions. For convenience, for $\alpha(Z)$ we have implemented a step function centered at $Z_0$ and of lateral extension $\sigma$:[30]

$$\alpha(Z) = \left(1 + exp(-(Z - Z_0)/\sigma)\right)^{-1} \quad (3)$$

The previous model, initially introduced in Ref.[30] shares similarities with those used by Aberkane et al.[62], Kim et al.[19,63] or Priftis[16,29] for characterizing the titration of charged colloidal species in different contexts.

# 3 – Results and Discussion

## 3.1 – Nanoparticle based coacervates and complexes

Fig. 2a displays vials containing PDADMAC/CeO$_2$@PAA mixed dispersions at $Z_{+/-}$ (upper panel) and $Z_{-/+}$ (lower panel) charge ratios ranging from 0.1 to 10, prepared through Type I and Type II direct mixing protocols, respectively. Visual inspection of these vials reveals turbid dispersions, indicating the formation of micron-sized structures resulting from the association between polymers and NPs. Fig. 2b display images of PDADMAD/PAA and PDADMAC/CeO$_2$@PAA mixed dispersions prepared at charge stoichiometry after centrifugation at 4000 rpm for 10 minutes. In both cases, a phase separation is observed, liquid-liquid for the polymer/polymer and liquid-solid for the polymer-NP systems. The turbidity observed in Fig. 2a at $Z_{+/-} \sim Z_{-/+} \sim 1$ is therefore attributed to a complex precipitate. The PDADMAC/CeO$_2$@PAA samples were also observed under phase contrast microscopy (Fig. 2c), showing for Z = 0.7 the presence of spherical micron-sized particles (average diameter 1.3 µm), and for $Z$ = 1 large aggregates in the 20-40 µm range. Aggregation is found to result from the association of the micron-sized particles observed at $Z$ = 0.7, likely originating from screened electrostatic interaction. Furthermore, the submicron aggregates in Fig. 2c do not reorganize over time, for example by coalescence, which would confirm the conclusion that they are solid precipitates. Far from stoichiometry, that is for $Z$ < 0.5, dispersions are less turbid, and their scattering properties is attributed to the formation of PECs.[30] In this work, we will adopt the term PEC to refer to particle and polymer complexes as well, in reference with the terminology used for polymers.

In the case of the PTEA-*b*-PAm block copolymer comprising a cationic polyelectrolyte and a neutral block, the behavior of the mixed dispersions containing γ-Fe$_2$O$_3$ NPs deviates from the previously described results. Throughout the $Z_{+/-}$-range, the dispersions remain single-phase, and the coacervation/precipitation transition is replaced by the formation of γ-Fe$_2$O$_3$ aggregates at the nanometer range. The cryo-TEM images in Fig. 2d describe the core of these mixed aggregates, revealing aggregation numbers of the order of a few units, median diameter of 50 ± 10 nm, and NPs in close contact with one another. In the figure, one observes a 1-2 nm distance between particle surfaces in the cores, suggesting the presence of polymers in these interspaces.[36] Electrophoretic mobility measurements on PTEA-*b*-PAm/γ-Fe$_2$O$_3$@PAA indicate that the mixed aggregates are neutral, further confirming the presence of a neutral polymer layer on the surface.

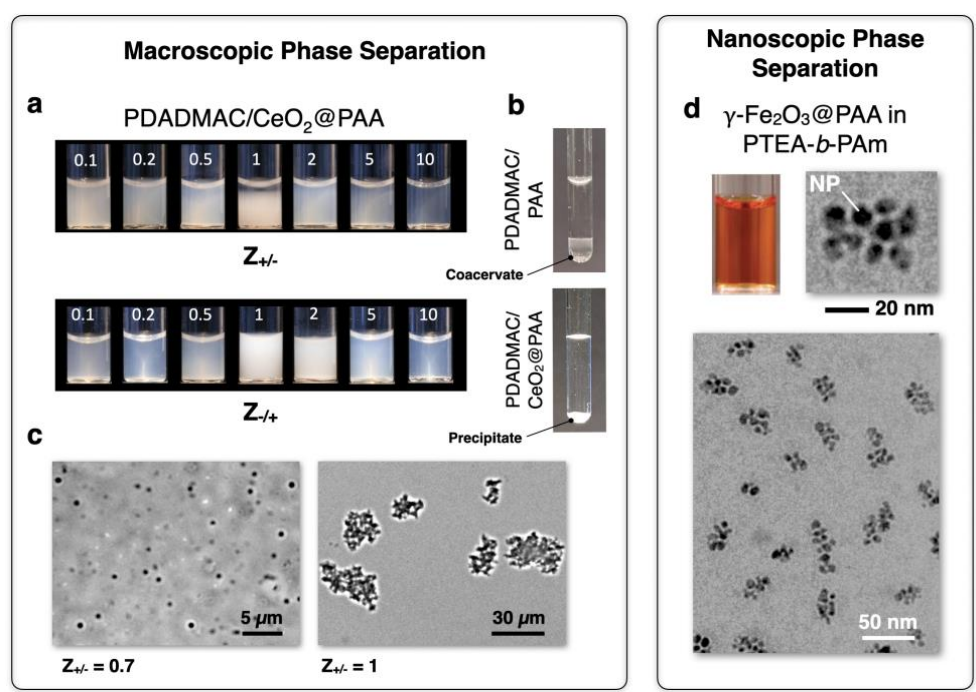

**Figure 2 :** a) Images of vials containing PDADMAC/CeO$_2$@PAA dispersions at charge ratios 0.1 to 10 following Type I and Type II mixing. b) PADADMAC/PAA (upper panel) and PDADMAC/CeO$_2$@PAA (lower panel) mixed dispersions at charge stoichiometry after a 4000 rpm centrifugation. c) Phase contrast microscopy images of the dispersions in a) for $Z_{+/-}$ = 0.7 (left panel) and $Z_{-/+}$ = 1 (right panel). The figure obtained under charge-matched conditions shows large aggregates of complex precipitate. d) Cryo-TEM image of PTEA-*b*-PAm/γ-Fe$_2$O$_3$@PAA resulting from Type I mixing. Insets : image of a vial containing a γ-Fe$_2$O$_3$@PAA dispersion (left); close-up view of an aggregate made from 10 γ-Fe$_2$O$_3$ NPs.

## 3.2 – Isothermal titration calorimetry
### 3.2.1 – Polymers (PDADMAC) and coated cerium oxide nanoparticles (CeO$_2$@PAA)

We first give an overview of results obtained for the polymer system, PDADMAC/PAA, serving as a reference for the subsequent data.[30] In Fig. 3a and Fig. 3b, binding isotherms for PDADMAC and PAA are depicted, acquired through Type I and Type II titrations at concentrations of 10/1 mM. For Type I experiments, enthalpy curves exhibit a sigmoidal decrease with increasing charge ratio $Z_{+/-}$. On charge neutralization, enthalpy reaches a shallow minimum at 0.1 kJ mol$^{-1}$ before a slight rise, followed by a second decrease beyond $Z_{+/-}$ = 2, indicating that the endothermic reaction is completed. In Type II experiments, we observe analogous endothermic behavior, marked by a distinct peak around charge-matched conditions ($Z_{-/+} \sim$ 1), with enthalpy approaching zero for ratios around 2.

Figs. 3c and 3d display the binding isotherms obtained from the titration of PDADMAC and CeO$_2$@PAA using the Type I and Type II mixing, respectively. Experiments were performed at concentrations 5/0.5 and 10/1 mM under the same conditions as above. In Type I experiments, reaction enthalpies are initially positive at low charge ratios, indicating again an endothermic process. Subsequently, $\Delta H(Z_{+/-})$ decreases sharply and becomes negative, with the minimum at -4 kJ mol$^{-1}$ coinciding precisely with charge stoichiometry. For $Z_{+/-} > 1$, the data rises and tends towards 0, suggesting completion of titration. For CeO$_2$@PAA NPs injected in PDADMAC polymer (Type II), two processes are again visible in the binding isotherms, however both are endothermic. These two sets of binding isotherms, those from Fig. 3a-3b and from Fig 3c-3d exhibit striking similarities, particularly the presence of two consecutive thermodynamic reactions. The initial process at low charge ratio is endothermic for both, whereas the second occurring around charge-matched conditions depends on the order of titration, exothermic for Type I and endothermic for Type II titration.

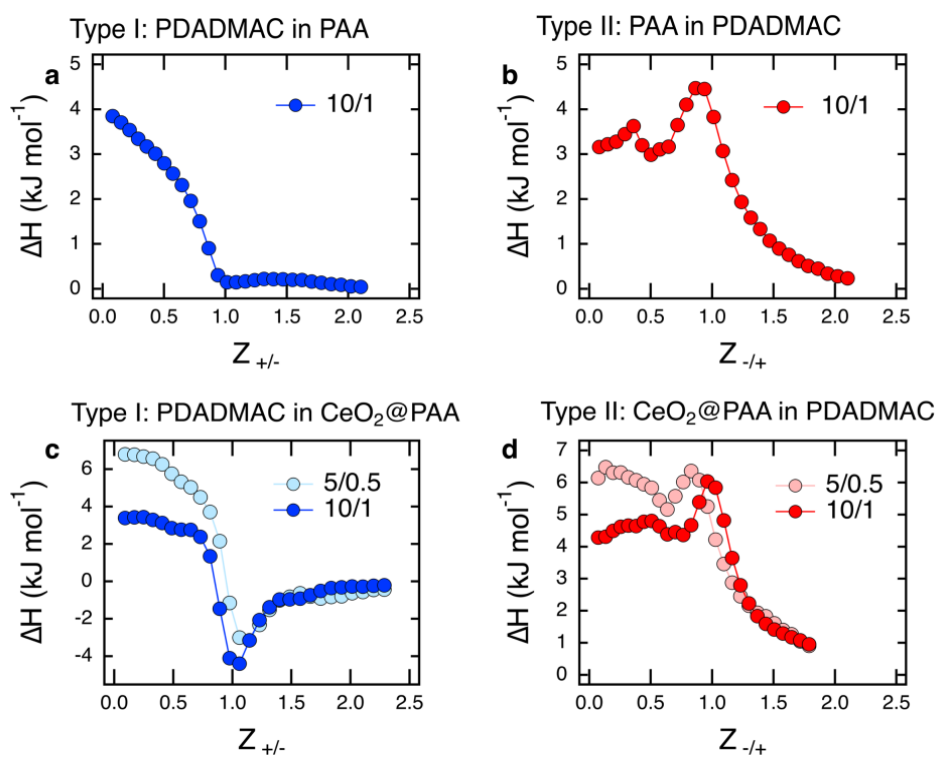

**Figure 3**: Binding isotherms of PDADMAC/PAA and PDADMAC/CeO$_2$@PAA dispersions obtained through Type sI (a,b) and Type II (c,d) mixing modes. The reaction enthalpies are plotted as a function of the charge ratios $Z_{+/-} = [+]/[-]$ and $Z_{-/+} = [-]/[+]$. The legends 5/0.5 and 10/1 indicate the molar concentrations of electrostatic charges (expressed in mM) for the titrating and titrated dispersions. The data in a) and c) are from Ref.[30].

**3.2.2 – Block copolymers (PTEA-*b*-PAm) and coated iron oxide nanoparticles (γ-Fe$_2$O$_3$@PAA)**

As in the previous section, we first illustrate the binding curves of the polymers alone, PTEA-*b*-PAm and PAA using Type I (Fig. 4a) and Type II (Fig. 4b) titrations. For experiments carried out at concentrations 10/1, 20/2 and 40/4 mM, positive enthalpy values starting around 3 kJ mol$^{-1}$ are observed at titration onset. Out of these results, only the 10/1-dataset shows a non-monotonic behavior, with a local minimum at $Z_{+/-}$ = 0.75. For titrations of Type II, as PAA is added to block copolymers, the reaction enthalpies are slightly lower, and also fairly well superimposed. This time however, the thermodynamic data exhibit a marked maximum of charge stoichiometry $Z_{-/+}$ = 1. Figs. 4c and 4d display the binding isotherms for block copolymers (PTEA-*b*-PAm) and coated iron oxide NPs (γ-Fe$_2$O$_3$@PAA). In Type I experiments, initial positive reaction enthalpies at low charge ratios evidence an endothermic process at the titration onset. Following this, $\Delta H(Z_{+/-})$ undergoes a sharp decrease and becomes negative, reaching its minimum at charge stoichiometry. With increasing $Z_{+/-}$, the data exhibits an upward trend, approaching 0, indicating that the titration is terminated. It should be noted that at 40/4 mM, the isotherm appears smeared and does not show at first clear evidence of exothermic heat exchange. The data in Figs. 4c exhibit strong similarities with those obtained with cationic PDADMAC homopolymers and cerium oxide NPs. For the Type II titration data in Fig. 4d, we also find similar patterns, *i.e.* an endothermic response with a peak centered around charge-matched conditions.

Fig. 4d reveals however a notable difference: for charge ratios $Z_{-/+}$ smaller than 0.5, we observe a distinct increase in the thermodynamic signal, starting from zero, a trend not observed in any of the other titration curves. This result is reminiscent of the critical aggregation concentration phenomenon observed in ITC with dodecyltrimethylammonium bromide surfactant micelles[24] and in light scattering with metal oxide NPs.[54] Notably, it is unique to Type II titration involving copolymers. Overall, the thermodynamic signatures of all four pairs of opposing colloidal charge systems, including PDADMAC/PAA, PDADMAC/CeO$_2$@PAA, and PTEA-*b*-PAm/PAA and PTEA-*b*-PAm/γ-Fe$_2$O$_3$@PAA bear strong similarities, indicating a behavior unaffected by the nature of the complexing species, whether macromolecular or colloidal.

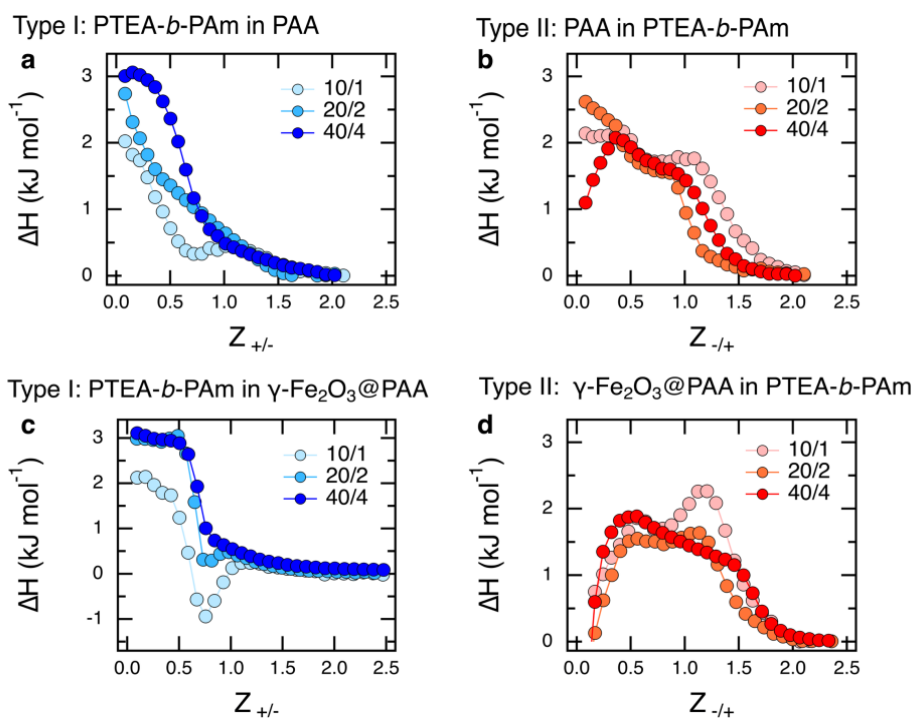

**Figure 4**: Binding enthalpies measured by isothermal titration calorimetry between oppositely charged (a,b) polymers (PTEA-*b*-PAm, PAA) and (c,d) polymers and NPs (PTEA-*b*-PAm, γ-Fe$_2$O$_3$@PAA). The diagrams on the left-hand side (a,c) show data for Type I titration, the charge ratio being $Z_{+/-} = [+]/[-]$. Diagrams on the right-hand side display Type II titration with charge ratio $Z_{-/+} = [-]/[+]$. The values indicated in the legends (10/1, 20/2 and 40/4) are the charge molar concentrations of the titrating and titrated dispersions, respectively.

## 3.3 - Thermodynamic versus structural titration

Figs. 5a, 5b and 5c compare the Type I binding enthalpy $\Delta H(Z_{+/-})$ obtained on the system PDADMAC/CeO$_2$@PAA with data acquired through static and dynamic light scattering performed under the same titration conditions. The aim is to establish correlations between the heat exchange processes and associated structural changes. The thermodynamic and structural results shown here are for concentrations 10 mM for the titrant and 1 mM for the titrate. In Fig. 5b, the progressive increase in scattered intensity $I_S(Z_{+/-})$ upon PDADMAC addition is indicative

of the formation of PECs, characterized here by a hydrodynamic diameter $D_H$ = 120 ± 10 nm (Fig. 5c). This first regime continues until the critical charge ratio $Z_{+/-}^C$ = 0.64, at which ratio $D_H(Z_{+/-})$ grows sharply into the micron range. This growth is associated with the transition to the complex precipitate identified by optical microscopy (Fig. 2c). Beyond $Z_{+/-}^C$, the scattered intensity shows a non-monotonic behavior associated with high absorbance and/or rapid sedimentation/coalescence of the precipitate. It can be pointed out that precipitation takes place precisely to the onset of the exothermic reaction, where the enthalpy decreases rapidly and becomes negative. In Type II titration, similar patterns are observed for the binding enthalpy $\Delta H(Z_{-/+})$, scattering intensity $I_S(Z_{-/+})$, and hydrodynamic diameter $D_H(Z_{-/+})$, as depicted in Figs. 5d, 5e and 5f, respectively. Here, the transition takes place at a slightly higher charge ratio, $Z_{-/+}^C$ = 0.75. The most important difference, however, is that the transition from PECs to complex precipitate of PDADMAC/CeO$_2$@PAA is an endothermic process, with peak enthalpy peaking at around 6 kJ mol$^{-1}$ and $Z_{-/+}$ = 1. In parallel, $\zeta$-potential measurements evidenced that the 120 nm aggregates formed at low charge ratio were negatively charged, with a $\zeta$-potential of $-44\ mV$. Upon PDADMAC addition, the $\zeta$-potential increased progressively and eventually became positive above charge neutrality. Conversely, Type II aggregates are initially positive ($\zeta$ = + 40 mV), with a $\zeta$-potential decreasing progressively toward 0 mV (Supporting Information S5).

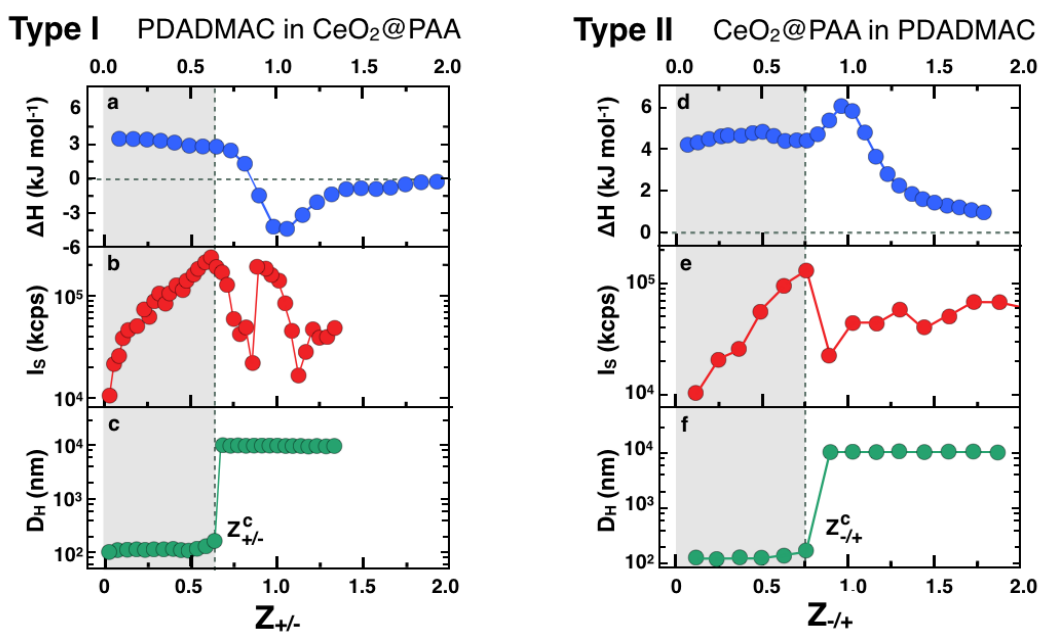

**Figure 5**: a) Binding enthalpy $\Delta H(Z_{+/-})$, b) scattering intensity $I_S(Z_{+/-})$ and c) hydrodynamic diameter $D_H(Z_{+/-})$ determined from Type I titration experiments performed on PDADMAC and CeO$_2$@PAA. The cationic and anionic charge concentrations are [+] = 10 mM and [−] = 1 mM. d,e,f) Same as a,b,c) for Type II titration. The critical charge ratios $Z_{+/-}^C$ = 0.64 and $Z_{-/+}^C$ = 0.75 indicate the transition between electrostatic complexes (indicated by the area in grey) and precipitation phase. In Figs. 5c and 5f, the hydrodynamic diameter is set at 10$^4$ nm in the precipitation phase.

A methodology comparable to that used in Fig. 5 was employed for PTEA-*b*-PAm/PAA and PTEA-*b*-PAm/$\gamma$-Fe$_2$O$_3$@PAA, as displayed in Supporting Information S6 and Fig. 6 respectively. In the Type I titration (Fig. 6a), where the copolymer is gradually introduced to a particle dispersion, at respective charge concentrations [+] = 20 mM and [−] = 2 mM, the scattered intensity $I_S(Z_{+/-})$ exhibits a rapid growth with increasing charge ratio. It then reaches a plateau when the binding enthalpy undergoes a sharp decrease. At the same time, the hydrodynamic diameter $D_H(Z_{+/-})$ decreases slightly from 150 nm before stabilizing above $Z_{+/-}$ = 0.5 and maintaining a constant value around 110 ± 10 nm. For Type II titration (Fig. 6b), light scattering data exhibit lower scattering intensities and hydrodynamic diameters compared to Type I data, with $D_H$-values in the range of 50-60 nm. Notably, both $I_S(Z_{-/+})$ and $D_H(Z_{-/+})$ display a maximum around Z = 1.3 ± 0.1, coinciding with the decrease in binding enthalpy. The main conclusion drawn from the combination of structural and thermodynamic titrations is that, akin to coacervation, the formation of aggregated nanostructures occurs in two stages. During the first injection, aggregates spontaneously form but undergo reorganization as the charge stoichiometry nears. In parallel $\zeta$-potential measurements consistently show values close to 0 across the entire range of charge ratios, a scenario markedly different from that observed with homopolymers.

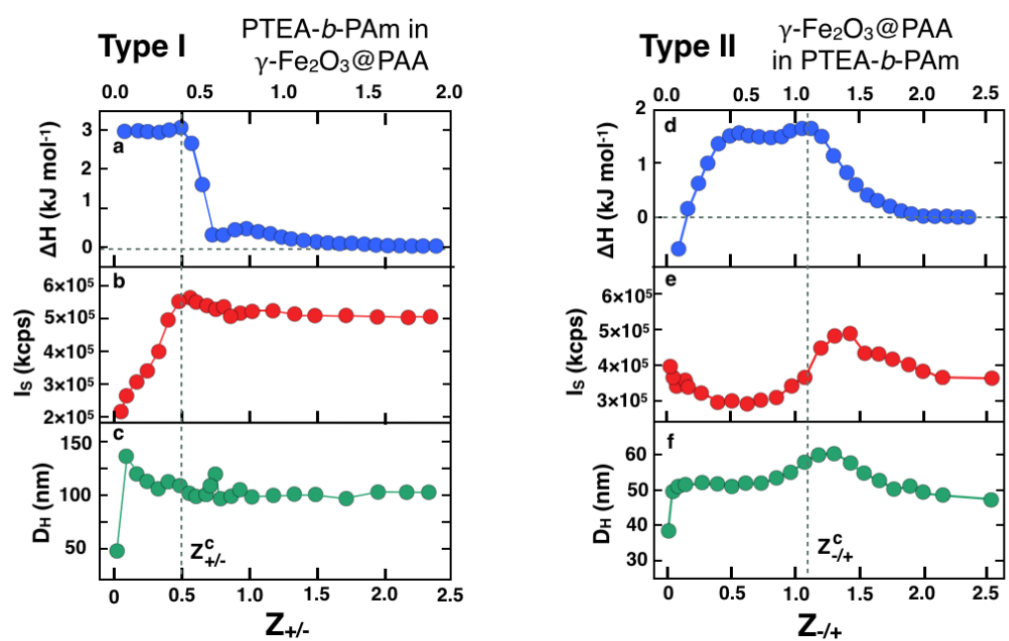

**Figure 6**: a) Binding enthalpy $\Delta H(Z_{+/-})$, b) scattering intensity $I_S(Z_{+/-})$ and c) hydrodynamic diameter $D_H(Z_{+/-})$ determined from Type I titration experiments performed on PTEA-*b*-PAm and $\gamma$-Fe$_2$O$_3$@PAA. The cationic and anionic charge concentrations are [+] = 20 mM and [−] = 2 mM. d,e,f) Same as a,b,c) for Type II titration.

### 3.4 – ITC titration curve analysis
Fig. 7 displays the ITC data for PDADMAC/CeO$_2$@PAA and PTEA-*b*-PAm/$\gamma$-Fe$_2$O$_3$@PAA together with the fitting curves obtained from the modified version of the Multiple Non-Interacting Sites model (MNIS, Eq. 2 and 3). In the MNIS model, the adjustable parameters are the binding enthalpy $\Delta H_b^{P,S}$, the binding constant $K_b^{P,S}$ and the reaction stoichiometry $n^{P,S}$, each quantity being indexed with the letter $P$ or $S$ in reference to the primary or secondary process. To constrain the number of adjustable variables, the step function $\alpha(Z)$ in Eq. 3 is set up such that $Z_0 = n^{P,S}$ and $\sigma = 0.08$ for each set of curves, a choice that resulted in satisfactory fits across the entire Z-range examined. In Fig. 7a-j, the experimental data obtained for PDADMAC/CeO$_2$@PAA (at concentrations 10/1 and 20/2 mM) and PTEA-*b*-PAm/γ-Fe$_2$O$_3$@PAA (at concentrations 10/1, 20/2 and 40/4 mM) are represented by dots, while the solid blue and green lines illustrate the contributions from the primary and secondary processes, respectively. The solid red line represents the combined contributions, demonstrating excellent agreement with the data. It is worth noting that the single-reaction MNIS scheme, *via* Eq.1 was also evaluated for completeness but could not account for the ITC data. Data for the PTEA-*b*-PAm/PAA system can be found in Supporting Information S7. To facilitate comparison across the various systems investigated, PDADMAC/PAA and PTEA-*b*-PAm/PAA, PDADMAC/CeO$_2$@PAA and PTEA-*b*-PAm/γ-Fe$_2$O$_3$@PAA, the thermodynamic data collected were organized in tabular format in the Supporting Information S8 and represented in histogram form in Fig. 8 and Supporting Information S9. This presentation allows for a more comprehensive assessment of the thermodynamic variables underlying these reactions.

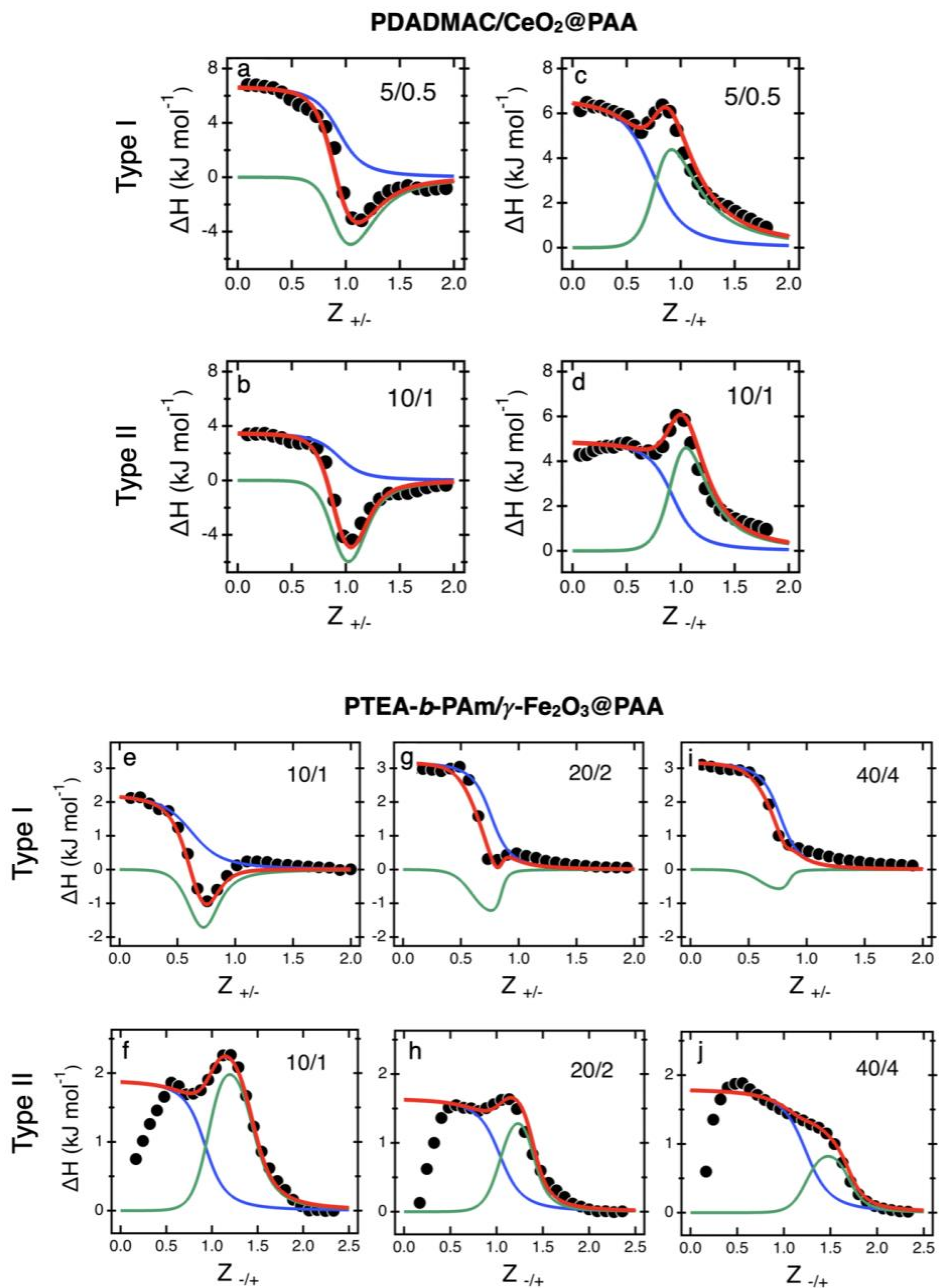

**Figure 7**: a-d) Binding isotherms for PDADMAC/CeO$_2$@PAA titration curves at different concentrations and for Type I (a,c) and Type II (b,d) experiments. e-i) Same as in a-d) for the PTEA-b-PAm/γ-Fe$_2$O$_3$@PAA system. The solid blue and green lines illustrate the contributions of the primary and secondary processes, respectively, while the solid red line represents their combined effects. The continuous lines through the data points are from Eq. 2, with parameters indicated in Tables S3 and S4.

Referring to the data presented in Tables S1-S4, we first examine the evolution of stoichiometry, represented by $n^P$ and $n^S$, as well as the binding constants $K^P$ and $K^S$. Across all examined systems, the values of $n^P$ and $n^S$ are consistently near unity, although $n^P < n^S$ is systematically observed, confirming the sequential nature of the thermodynamic titration. The binding constants range between $10^2$ and $10^4$ M$^{-1}$, in line with literature results for similar systems.[19,24,29,30,63,64] The $K_b^P$-ratios between Type I and Type II titrations are generally higher for NPs compared to polymers, typically by a factor of 3. Additionally, the low $K_b^P$ and $K_b^S$ values for the PTEA-*b*-PAm/PAA system at 20/2 mM are attributed to the difficulty of fitting the reaction enthalpy with the MNIS model, as shown in Fig. S6c. To analyze the changes in enthalpy and entropy, we focus on the binding enthalpy $\Delta H_b^{P,S}$ and the entropic contribution to the total free energy $-T\Delta S^{P,S}$ in Fig. 8. In regard to the primary process (Fig. 8a-d), it is observed that the $\Delta H_b^P$ are positive, ranging between +2 and +6 kJ mol$^{-1}$, indicating that the PEC formation at titration onset proceeds *via* an endothermic reaction. Another significant observation regarding the primary process is that the entropy contribution $-T\Delta S^P$ ranges between +20 and +25 kJ mol$^{-1}$, and consistently surpasses the enthalpy in all instances. This supports the findings from various ITC experiments suggesting that the charge association reaction is primarily driven by an entropic process related to the condensed counterion release.[8,16,30,48,65] The notable difference between the primary and secondary processes can be seen in Figs. 8e-h, where the binding enthalpy values change sign depending on the mixing mode, negative for Type I titrations and positive for Type II titrations. This phenomenon is fairly general and applies to homopolymers, copolymers and nanoparticles alike. We also note that the exothermic effect is stronger for homopolymers than for copolymers, with a delta $\Delta H_b^S$ of up to -10 kJ mol$^{-1}$ for PDADMAC/CeO$_2$@PAA. From a thermodynamic perspective, these results indicate that there is no reciprocity between positive and negative charges concerning the mixing order. This is a surprising and, to our knowledge, as yet unresolved result in the titration of colloids and polymers with opposite charges. It may arise from the structures and charge distributions of the PECs that are spontaneously assembled during the primary process and which undergo transformations towards coacervate or complex precipitate with the continuous addition of oppositely charged species.

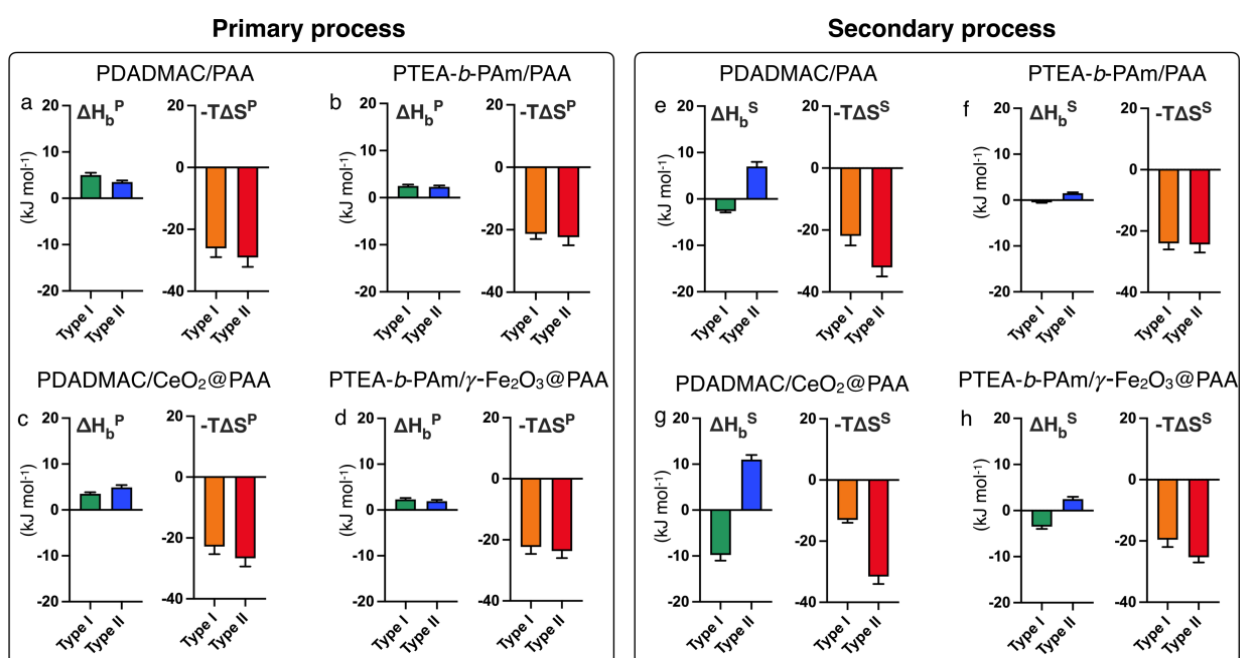

**Figure 8**: a-d) Binding enthalpy $\Delta H_b^P$ and entropic contribution to the total free energy $-T\Delta S^P$ extracted from ITC thermograms analyses and related to the primary process for the following

colloidal systems: a) PDADMAC/PAA, b) PTEA-*b*-PAm/PAA, c) PDADMAC/CeO$_2$@PAA and d) PTEA-*b*-PAm/γ-Fe$_2$O$_3$@PAA. e-h) Same as for a-d) for $\Delta H_b^S$ and $-T\Delta S^S$ associated with the secondary process. Note that in the histograms in e-h), the binding enthalpies change sign according to the mixing mode. All data presented correspond to molar charge concentrations of 10 mM for the titrant and 1mM for the titrate. The error bars in the histograms are derived from the fitting procedure using Eq. 1.

# IV – Conclusion

In this study, we investigate the complexation between oppositely charged polymers and sub 10 nm metal oxide NPs using isothermal titration calorimetry, light scattering, electrophoresis and microscopy. By combining different charged species, including polyelectrolytes, charged neutral block copolymers, and poly(acrylic acid) coated metal oxide nanoparticles we aimed to explore a wide variety of complexation schemes beyond those typically found on polymers or proteins. A total of 4 systems are identified, namely polymer/polymer, polymer/NP, copolymer/polymer and copolymer/NP. Mixing oppositely charge species have been implemented using direct mixing and titration. In this work, we also focus on mixing order, an issue that has received little attention in the literature, and we highlight its significance in the context of reaction thermodynamics. The results obtained by titration from all four above mentioned dispersions were found to follow common behaviors as a function of charge ratios. In particular, the ITC and light scattering data reveal in all examples examined a two-step titration process. The primary step involves the formation of charged PECs with a size of around 100 nm. Subsequently, the transition from charged PECs to coacervate droplets (for PDADMAC/PAA) or complex precipitate (for PDADMAC/CeO$_2$@PAA) occurs around charge-matched conditions. The use of a modified non-interacting multi-site (MNIS) model enables quantitative analysis of ITC data, highlighting the prevalence of the entropic contribution over the enthalpic one to the reaction free energy. The data highlight also the non-reciprocity of the mixing mode in terms of the reaction enthalpy associated with coacervation and precipitation. Similarly, the reorganization of PTEA-*b*-PAm/γ-Fe$_2$O$_3$@PAA aggregates at the charge stoichiometry shows enthalpies of opposite sign depending on the mixing mode. These transitions are characterized by exothermic profiles for Type I titration (*i.e.* addition of cationic species to anionic ones) and endothermic profiles for Type II titration (i.e. addition of anionic species to cationic ones). These results could be associated with the fact that PECs obtained by titration exhibit structures as well as charge distributions that depend on the order of mixing, as exemplified by light scattering and electrophoretic mobility. Such a result has not been systematically studied in the realm of coacervation/precipitation, and would require further experimental and theoretical studies. While ITC remains a valuable technique for retrieving thermodynamic parameters, our study finally underscores the importance of performing parallel structural, electrophoresis and optical microscopy experiments under the same conditions of concentration, pH and injection, an approach that enables a reliable correlation to be established between thermodynamics, charge and structure.

# Associated content
**Supporting Information**

*Tuesday, October 29, 24*17

Transmission electron microscopy images of cerium and iron oxide nanoparticles (S1) – Relationship between $CeO_2$@PAA and $\gamma$-$Fe_2O_3$@PAA nanoparticle weight and charge concentrations (S2) – Calibration of the Microcal VP-ITC calorimeter using $CaCl_2$, EDTA in MES buffer (S3) – Examples of dilution curves obtained from isothermal titration calorimetry for the 5 systems investigated (S4) – Electrophoretic mobility obtained from PDADMAC/$CeO_2$@PAA titration (S5) – PTEA-b-PAm/PAA thermodynamic and structure titrations (S46) – ITC titration curve analysis for PTEA-b-Pam/PAA (S7) – Results of fitting titration curves with the MNIS model (S8) – Effect of titrant and titrate concentration on thermodynamic parameters extracted from MNIS model fitting (S9)

# Author Information
**Corresponding Author**
Jean-François Berret – Université Paris Cité, CNRS, Matière et systèmes complexes, 75013 Paris, France ; orcid.org/0000-0001-5458-8653 ; email : jean-francois.berret@u-paris.f

**Author**
Letícia Vitorazi – Laboratório de Polímeros, Nanomateriais e Química Supramolecular, EEIMVR, Universidade Federal Fluminense, Avenida dos Trabalhadores, 420, Volta Redonda RJ, CEP 27225-125, Brazil and Programa de Pós-Graduação em Engenharia Metalúrgica, EEIMVR, Universidade Federal Fluminense, Avenida dos Trabalhadores, 420, Volta Redonda RJ, CEP 27225-125, Brazil; orcid.org/0000-0002-6626-9016; email: leticiavitorazi@id.uff.br

**Author Contributions**
L.V.: methodology, investigation, visualization, formal analysis, writing-first draft.
J.-F. B.: conceptualization, methodology, validation, writing-final draft and review, editing, project administration, supervision and funding acquisition.
**Notes**
The authors declare no competing financial interest.

# Acknowledgment
We are grateful to J. Fresnais, F. Mousseau, E.K. Oikonomou for fruitful discussions and the PHENIX laboratory for support in light scattering and electrophoresis titration experiments. Agence Nationale de la Recherche (ANR) and Commissariat à l'Investissement d'Avenir are gratefully acknowledged for their financial support of this work through Labex SEAM (Science and Engineering for Advanced Materials and devices) ANR-10-LABX-0096 et ANR-18-IDEX-0001. This research was supported in part by the ANR under the contracts ANR-13-BS08-0015 (PANORAMA), ANR-12-CHEX-0011 (PULMONANO), ANR-15-CE18-0024-01 (ICONS), ANR-17-CE09-0017 (AlveolusMimics) and by Solvay. L.V. also thanks the CNPq (Conselho Nacional de Desenvolvimento Científico e Tecnológico), Proc. N$^0$ 210694/2013-0 in Brazil.

# TOC figure

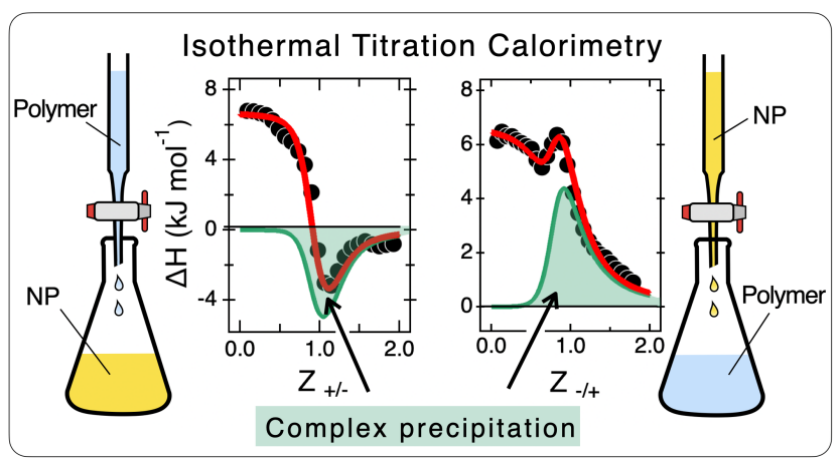

Mixing order induced asymmetric heat exchange associated to complex precipitation during thermodynamic polymer-nanoparticle titration